\documentclass[twocolumn]{aastex62}

\usepackage{bm}
\usepackage{natbib,bm,multirow,tabularx}
\shortauthors{Chung et al.}

\newcommand{\te}{t_{\rm E}}

\newcommand{\tzerone}{t_{01}}
\newcommand{\tzerotwo}{t_{02}}
\newcommand{\uzero}{u_{0}}
\newcommand{\uzerone}{u_{01}}
\newcommand{\uzerotwo}{u_{02}}

\newcommand{\pie}{\pi_{\rm E}}
\newcommand{\pien}{\pi_{\rm E,N}}
\newcommand{\piee}{\pi_{\rm E,E}}
\newcommand{\bpie}{\bm{\pi}_{\rm E}}

\newcommand{\thetae}{\theta_{\rm E}}
\newcommand{\thetaemin}{\theta_{\rm E,min}}
\newcommand{\thetas}{\theta_{\rm \star}}

\newcommand{\murel}{\mu_{\rm rel}}

\newcommand{\dl}{D_{\rm L}}
\newcommand{\ds}{D_{\rm S}}
\newcommand{\delcs}{\Delta \chi^{2}}
\newcommand{\thisevent}{KMT-2022-BLG-0086}

\begin{document}

\title{\thisevent: Another binary-lens binary-source microlensing event}

\correspondingauthor{Sun-Ju Chung}
\email{sjchung@kasi.re.kr}

\author{Sun-Ju Chung}
\affiliation{Korea Astronomy and Space Science Institute, 776 Daedeokdae-ro, Yuseong-Gu, Daejeon 34055, Republic of Korea}

\author{Kyu-Ha Hwang}
\affiliation{Korea Astronomy and Space Science Institute, 776 Daedeokdae-ro, Yuseong-Gu, Daejeon 34055, Republic of Korea}

\author{Jennifer C. Yee}
\affiliation{Center for Astrophysics $|$ Harvard \& Smithsonian, 60 Garden St., Cambridge, MA 02138, USA} 

\author{Andrew Gould}
\affiliation{Department of Astronomy, Ohio State University, 140 W. 18th Avenue, Columbus, OH 43210, USA}
\affiliation{Max-Planck-Institute for Astronomy, Königstuhl 17, D-69117 Heidelberg, Germany}

\author{Ian A. Bond}
\affiliation{Institute of Natural and Mathematical Science, Massey University, Auckland 0745, New Zealand}

\author{Hongjing Yang}
\affiliation{Department of Astronomy and Tsinghua Centre for Astrophysics, Tsinghua University, Beijing 100084, China}

\collaboration{(Leading Authors),}

\author{Michael D. Albrow}
\affiliation{Department of Physics and Astronomy, University of Canterbury, Private Bag, 4800 Christchurch, New Zealand}

\author{Youn Kil Jung}
\affiliation{Korea Astronomy and Space Science Institute, 776 Daedeokdae-ro, Yuseong-Gu, Daejeon 34055, Republic of Korea}
\affiliation{University of Science and Technology, Korea, (UST), 217 Gajeong-ro, Yuseong-gu, Daejeon 34113, Republic of Korea}

\author{Cheongho Han}
\affiliation{Department of Physics, Chungbuk National University, Cheongju 361-763, Republic of  Korea}

\author{Yoon-Hyun Ryu}
\affiliation{Korea Astronomy and Space Science Institute, 776 Daedeokdae-ro, Yuseong-Gu, Daejeon 34055, Republic of Korea}

\author{In-Gu Shin}
\affiliation{Center for Astrophysics $|$ Harvard \& Smithsonian, 60 Garden St., Cambridge, MA 02138, USA}

\author{Yossi Shvartzvald}
\affiliation{Department of Particle Physics and Astrophysics, Weizmann Institute of Science, Rehovot 76100, Israel}

\author{Weicheng Zang}
\affiliation{Center for Astrophysics $|$ Harvard \& Smithsonian, 60 Garden St., Cambridge, MA 02138, USA}

\author{Sang-Mok Cha}
\affiliation{Korea Astronomy and Space Science Institute, 776 Daedeokdae-ro, Yuseong-Gu, Daejeon 34055, Republic of Korea}
\affiliation{School of Space Research, Kyung Hee University, Giheung-gu, Yongin, Gyeonggi-do, 17104, Korea}

\author{Dong-Jin Kim}
\affiliation{Korea Astronomy and Space Science Institute, 776 Daedeokdae-ro, Yuseong-Gu, Daejeon 34055, Republic of Korea}

\author{Seung-Lee Kim}
\affiliation{Korea Astronomy and Space Science Institute, 776 Daedeokdae-ro, Yuseong-Gu, Daejeon 34055, Republic of Korea}

\author{Chung-Uk Lee}
\affiliation{Korea Astronomy and Space Science Institute, 776 Daedeokdae-ro, Yuseong-Gu, Daejeon 34055, Republic of Korea}

\author{Dong-Joo Lee}
\affiliation{Korea Astronomy and Space Science Institute, 776 Daedeokdae-ro, Yuseong-Gu, Daejeon 34055, Republic of Korea}

\author{Yongseok Lee}
\affiliation{Korea Astronomy and Space Science Institute, 776 Daedeokdae-ro, Yuseong-Gu, Daejeon 34055, Republic of Korea}
\affiliation{School of Space Research, Kyung Hee University, Giheung-gu, Yongin, Kyeonggi 17104, Republic of Korea}

\author{Byeong-Gon Park}
\affiliation{Korea Astronomy and Space Science Institute, 776 Daedeokdae-ro, Yuseong-Gu, Daejeon 34055, Republic of Korea}
\affiliation{University of Science and Technology, Korea, (UST), 217 Gajeong-ro, Yuseong-gu, Daejeon 34113, Republic of Korea}

\author{Richard W. Pogge}
\affiliation{Department of Astronomy, Ohio State University, 140 W. 18th Avenue, Columbus, OH 43210, USA}
\collaboration{(The KMTNet collaboration),}

\author{Fumio Abe}
\affiliation{Institute for Space-Earth Environmental Research, Nagoya University, Nagoya 464-8601, Japan}

\author{David P. Bennett}
\affiliation{Code 667, NASA Goddard Space Flight Center, Greenbelt, MD 20771, USA}
\affiliation{Department of Astronomy, University of Maryland, College Park, MD 20742, USA}

\author{Aparna Bhattacharya}
\affiliation{Code 667, NASA Goddard Space Flight Center, Greenbelt, MD 20771, USA}
\affiliation{Department of Astronomy, University of Maryland, College Park, MD 20742, USA}

\author{Akihiko Fukui}
\affiliation{Department of Earth and Planetary Science, Graduate School of Science, The University of Tokyo, 7-3-1 Hongo, Bunkyo-ku, Tokyo 113-0033, Japan}
\affiliation{Instituto de Astrofísica de Canarias, Vía Láctea s/n, 38205 La Laguna, Tenerife, Spain}

\author{Ryusei Hamada}
\affiliation{Department of Earth and Space Science, Graduate School of Science, Osaka University, Toyonaka, Osaka 560-0043, Japan}

\author{Yuki Hirao}
\affiliation{Institute of Astronomy, Graduate School of Science, The University of Tokyo, 2-21-1 Osawa, Mitaka, Tokyo 181-0015, Japan}

\author{Stela Ishitani Silva}
\affiliation{Department of Physics, The Catholic University of America, Washington, DC 20064, USA}
\affiliation{Code 667, NASA Goddard Space Flight Center, Greenbelt, MD 20771, USA}

\author{Naoki Koshimoto}
\affiliation{Department of Earth and Space Science, Graduate School of Science, Osaka University, Toyonaka, Osaka 560-0043, Japan}

\author{Shota Miyazaki}
\affiliation{Institute of Space and Astronautical Science, Japan Aerospace Exploration Agency, 3-1-1 Yoshinodai, Chuo, Sagamihara, Kanagawa 252-5210, Japan}

\author{Yasushi Muraki}
\affiliation{Institute for Space-Earth Environmental Research, Nagoya University, Nagoya 464-8601, Japan}

\author{Tutumi Nagai}
\affiliation{Department of Earth and Space Science, Graduate School of Science, Osaka University, Toyonaka, Osaka 560-0043, Japan}

\author{Kansuke Nunota}
\affiliation{Department of Earth and Space Science, Graduate School of Science, Osaka University, Toyonaka, Osaka 560-0043, Japan}

\author{Greg Olmschenk}
\affiliation{Code 667, NASA Goddard Space Flight Center, Greenbelt, MD 20771, USA}

\author{Clément Ranc}
\affiliation{Sorbonne Université, CNRS, UMR 7095, Institut d’Astrophysique de Paris, 98 bis bd Arago, 75014 Paris, France}

\author{Nicholas J. Rattenbury}
\affiliation{Department of Physics, University of Auckland, Private Bag 92019, Auckland, New Zealand}

\author{Yuki Satoh}
\affiliation{Department of Earth and Space Science, Graduate School of Science, Osaka University, Toyonaka, Osaka 560-0043, Japan}

\author{Takahiro Sumi}
\affiliation{Department of Earth and Space Science, Graduate School of Science, Osaka University, Toyonaka, Osaka 560-0043, Japan}

\author{Daisuke Suzuki}
\affiliation{Department of Earth and Space Science, Graduate School of Science, Osaka University, Toyonaka, Osaka 560-0043, Japan}

\author{Sean K. Terry}
\affiliation{Code 667, NASA Goddard Space Flight Center, Greenbelt, MD 20771, USA}
\affiliation{Department of Astronomy, University of Maryland, College Park, MD 20742, USA}

\author{Paul J. Tristram}
\affiliation{University of Canterbury Mt. John Observatory, PO Box 56, Lake Tekapo 8770, New Zealand}

\author{Aikaterini Vandorou}
\affiliation{Code 667, NASA Goddard Space Flight Center, Greenbelt, MD 20771, USA}
\affiliation{Department of Astronomy, University of Maryland, College Park, MD 20742, USA}

\author{Hibiki Yama}
\affiliation{Department of Earth and Space Science, Graduate School of Science, Osaka University, Toyonaka, Osaka 560-0043, Japan}

\collaboration{(The MOA collaboration)}

\begin{abstract}
We present the analysis of a microlensing event \thisevent~of which the overall light curve is not described by a binary-lens single-source (2L1S) model, which suggests the existence of an extra lens or an extra source.
We found that the event is best explained by the binary-lens binary-source (2L2S) model, but the 2L2S model is only favored over the triple-lens single-source (3L1S) model by $\delcs \simeq 9$.
Although the event has noticeable anomalies around the peak of the light curve, they are not enough covered to constrain the angular Einstein radius $\thetae$, thus we only measure the minimum angular Einstein radius $\thetaemin$.
From the Bayesian analysis, it is found that that the binary lens system is a binary star with masses of $(m_1,m_2)=(0.46^{+0.35}_{-0.25}\, M_\odot, 0.75^{+0.67}_{-0.55}\, M_\odot)$ at a distance of $\dl=5.87^{+1.21}_{-1.79}$ kpc, while the triple lens system is a brown dwarf or a massive giant planet in a low-mass binary-star system with masses of $(m_1,m_2,m_3)=(0.43^{+0.41}_{-0.35}\, M_\odot, 0.056^{+0.055}_{-0.047}\, M_\odot, 20.84^{+20.20}_{-17.04}\, M_{\rm J})$ at a distance of $\dl=4.06^{+1.39}_{-3.28}$ kpc, indicating a disk lens system.
The 2L2S model yields the relative lens-source proper motion of $\murel \geqslant 4.6\, \rm mas\, yr^{-1}$ that is consistent with the Bayesian result, whereas the 3L1S model yields $\murel \geqslant 18.9\, \rm mas\, yr^{-1}$, which is more than three times larger than that of a typical disk object of $\sim 6\, \rm mas\, yr^{-1}$ and thus is not consistent with the Bayesian result.
This suggests that the event is likely caused by the binary-lens binary-source model.
\end{abstract}


\section{Introduction} \label{sec:intro}

Planets detected by microlensing are 4\% of over 5800\footnote{https://exoplanetarchive.ipac.caltech.edu/index.html} exoplanets discovered so far, but they are quite different from those detected by transit and radial velocity that detected  93\% of all the exoplanets.
All the microlensing planets are cold or ice planets widely separated from their host stars, while planets detected by the transit and radial velocity are mostly located close to their host stars.
Microlensing is also an unique method capable of probing the Galactic distribution of planets because it does not depend on the brightness of objects, but their mass.
Moreover, microlensing is sensitive to free-floating planets that are believed to have been ejected from their planetary systems (\citealt{Rasio&Ford1996, Kaib+2013, Malmberg+2011}).
Due to such advantages of microlensing, the space-based microlensing exoplanet survey \citep{Bennett&Rhie2002} was adopted for the \textit{Nancy Grace Roman Space Telescope} (\textit{Roman}) mission (\citealt{Spergel+2015, Akeson+2019}).
From the \textit{Roman}, it is expected that about 1400 bound planets with masses of $0.1-10^4\, M_{\oplus}$ \citep{Penny+2019} and about 250 free-floating planets with masses of $0.1-10^{3}\, M_{\oplus}$ \citep{Johnson+2020} would be discovered.

Planets are expected to be revealed in microlensing by a short-duration perturbation in an otherwise normal single-lens single-source (1L1S) event (\citealt{Mao&Paczynski1991}, \citealt{Gould&Loeb1992}).
This expectation has been borne out in the great majority of over 200 planets discovered by microlensing over two decades.
However, planets can also occur as short perturbations on binary-star microlensing events, where they can remain hidden or submerged in more complex light curves, until they are exhumed by careful analysis.
For example, OGLE-2007-BLG-349 \citep{Bennett+2016} had two severely degenerate solutions between a star hosting two planets and a planet in a binary star, with $\delcs=0.39$, but it turned out that it was caused by the planet in the binary star because the binary star model is consistent with the flux limits from \textit{HST}.
Including the planet of OGLE-2007-BLG-349, nine planets orbiting binary stars, OGLE-2013-BLG-0341 \citep{Gould+2014}, OGLE-2008-BLG-092 \citep{Poleski+2014}, OGLE-2016-BLG-0613 \citep{Han+2017}, OGLE-2006-BLG-284 \citep{Bennett+2020}, OGLE-2018-BLG-1700 \citep{Han+2020}, KMT-2020-BLG-041 (\citealt{Zang+2021}, \citealt{Zhang+2024}), OGLE-2023-BLG-0836 \citep{Han+2024b}, and KMT-2024-BLG-0404 (Han et al. submitted), have been discovered through microlensing so far.

However, the residuals from stellar-binary models are not necessarily due to a planet.
They could be due to a second source, such as OGLE-2018-BLG-0584, KMT-2018-BLG-2119 \citep{Han+2023}, KMT-2021-BLG-0284, KMT-2022-BLG-2480, or KMT-2024-BLG-0412 \citep{Han+2024a}, or possibly other effects including the microlens parallax and lens orbital motion effects, such as MACHO-97-BLG-41 (\citealt{Bennett+1999}, \citealt{Albrow+2000}, \citealt{Jung+2013}), OGLE-2013-BLG-0723  (\citealt{Udalski+2015}, \citealt{Han+2016}), KMT-2021-BLG-0322 \citep{Han+2021}.
Hence, careful work is required to distinguish among the possibilities.
The event \thisevent~also has a severe degeneracy between the binary-lens binary-source and the triple lens interpretations.
In this paper, we report the results of careful analyses for the event.

\section{Observations and data} \label{sec:obs}
The microlensing event \thisevent~occurred on a background source star at equatorial coordinates $\rm (RA, decl.)$=(17:33:57.68,-27:06:02.52), corresponding to the Galactic coordinates $(l, b)=(0 \fdg 19, 3 \fdg 15)$.

The event was first discovered by the Korea Microlensing Telescope Network (KMTNet; \citealt{Kim+2016}), which is an optimized system for detecting microlensing exoplanets.
KMTNet uses three identical 1.6 m telescopes with 4 deg$^2$ field of view (FOV) cameras that are globally distributed at the Cerro Tololo Inter-American Observatory in Chile (KMTC), the South African Astronomical Observatory (KMTS), and the Siding Spring Observatory in Australia (KMTA).
The KMTNet observations were carried out in the $I$ and $V$ bands.
The event lies in the KMT subprime field BLG15 with a cadence of $\Gamma \simeq 1\, \rm hr^{-1}$.

The Microlensing Observations in Astrophysics (MOA; \citealt{Bond+2001}) also independently found this event and it was designated as MOA-2022-BLG-130.
MOA uses a 1.8 m  telescope with $2.2\, \rm deg^{2}$ FOV at Mt. John Observatory in New Zealand.
The MOA observations were carried out in the MOA-Red band $R_{\rm MOA}$ that corresponds to the sum of the Cousins $R$ and $I$ bands.

KMTNet data were reduced by the pySIS photometry pipeline \citep{Albrow+2009} based on difference image analysis (DIA; \citealt{Alard&Lupton1998}).
For light curve modeling, the KMTNet data were re-reduced using tender-loving-care (TLC) PySIS pipeline that provides best quality data sets.
From the TLC photometry, we checked that one highly magnified data point at $\rm HJD-2450000\ (HJD^\prime) \simeq 9665.5$ is real. (see Figure 1).
In addition, KMTC $I$- and $V$-band images were reduced using the pyDIA code \citep{Albrow2017} in order to estimate the source color and construct the color-magnitude diagram (CMD) of stars around the source.
MOA data were reduced with the MOA's DIA photometry pipeline \citep{Bond+2001}.
According to the procedures of \citet{Yee+2012}, the errors of the photometric data obtained from each pipeline were rescaled to make them $\chi^2/{\rm dof} \rightarrow 1$, where dof represents the degrees of freedom.

\begin{figure}[t!]
\centering
\includegraphics[width=0.5\textwidth]{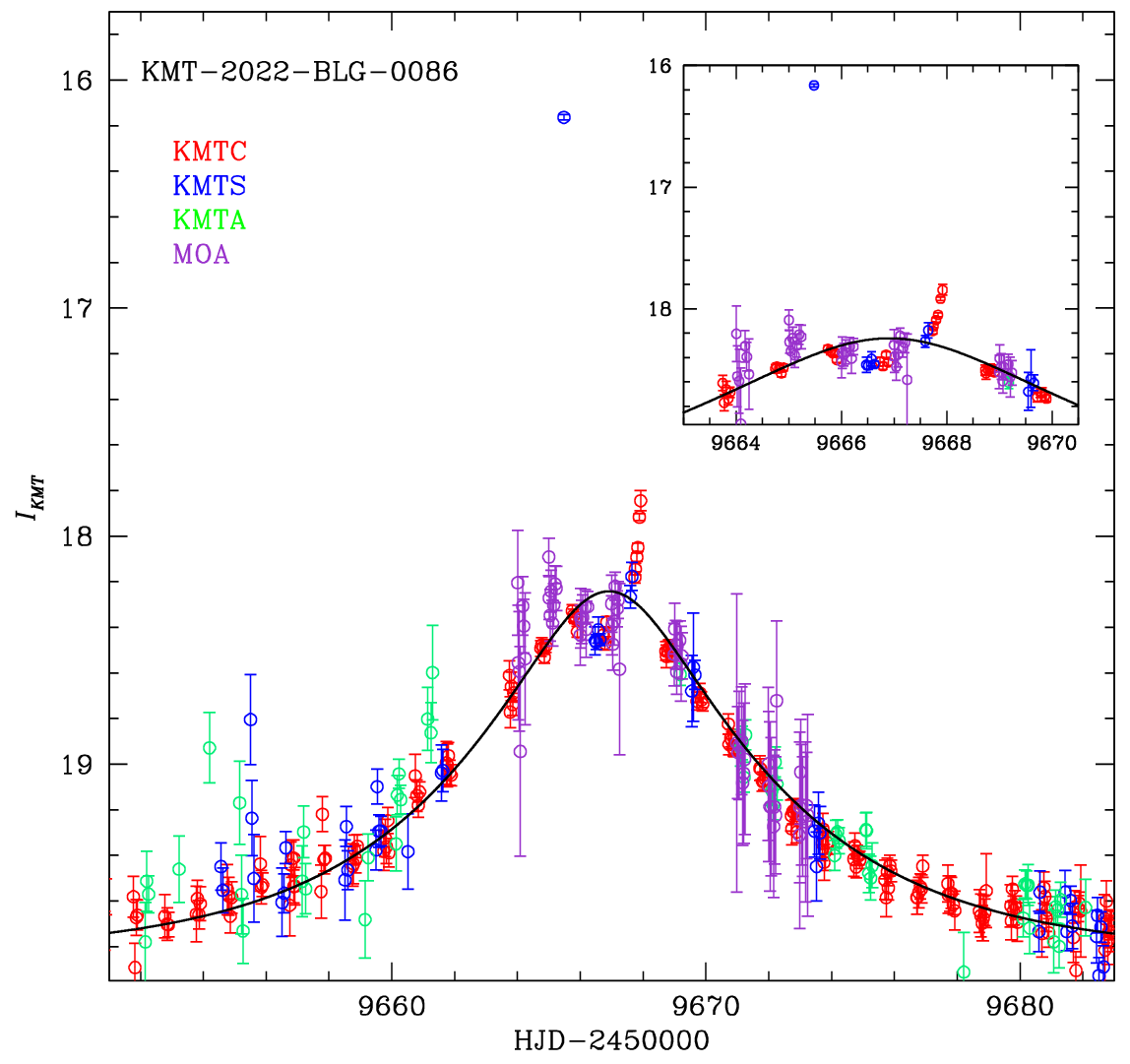}
\caption{Light curve of the microlensing event \thisevent.~The black curve is the single-lens single-source (1L1S) model fit.
\label{fig:1L1S}}
\end{figure}

\begin{deluxetable*}{lcccccc}
\tablewidth{0pt}
\tablecaption{Best-fit lensing parameters of two 2L1S models\label{tab-2L1S}}
\tablehead{
 Parameter                                    &&                    \multicolumn{2}{c}{Standard}                              &&          \multicolumn{2}{c}{Parallax+Orbital}        \\
 \cline{2-7}                                   &&                Model 1                    &                Model 2               &&              Model 1                    &        Model 2               
 }
\startdata
$\chi^2$                                       &&   $1676.65$                           &    $1763.34$                        &&    $1600.29$                        &    $1708.62$                     \\
$t_0$ (HJD$^\prime$)                   &&   $9666.8008 \pm 0.0385$   &  $9666.4049 \pm 0.0314$   &&   $9667.0999 \pm 0.0732$ &  $9666.2740 \pm 0.0457$ \\
$u_0$                                           &&  $0.0592 \pm 0.0025$          & $0.0772 \pm 0.0045$         &&   $0.0333 \pm 0.0058$        & $0.0660 \pm 0.0065$         \\
$\te$ (days)                                  &&  $19.1909 \pm 0.8423$        & $18.6452 \pm 0.8430$       &&   $36.6228 \pm 4.1230$      & $21.1662 \pm 1.3840$       \\
$s$                                               &&  $0.4684 \pm 0.0094$          & $0.4967 \pm 0.0163$         &&   $0.4335 \pm 0.0169$        & $0.4100 \pm 0.0165$         \\
$q$                                              &&  $0.3569 \pm 0.0276$          & $0.3827 \pm 0.0365$         &&   $0.1384 \pm 0.0377$         & $1.6245 \pm 0.1737$        \\
$\alpha$ (radians)                        &&  $2.4089 \pm 0.0175$          & $3.7325 \pm 0.0164$         &&   $2.5607 \pm 0.0352$         & $3.6935 \pm 0.0197$        \\
$\rho$                                         &&  $< 0.0005$                          & $< 0.003$                              &&   $< 0.002$                           & $< 0.001$         \\
$\pien$                                        &&          ...                                 &                ...                          &&  $9.9419 \pm 2.4518$          &  $9.9873 \pm 0.5678$        \\
$\piee$                                        &&          ...                                 &                ...                          &&  $-0.6201 \pm 0.4840$        &  $0.1050 \pm 0.5393$      \\
$ds/dt$ (yr$^{-1}$)                      &&          ...                                 &                ...                          &&   $3.3107 \pm 1.0847$         &   $-8.1957 \pm 0.8869$      \\
$d\alpha/dt$ (radians yr$^{-1}$) &&          ...                                 &                ...                          &&   $5.2053 \pm 2.1390$       &   $0.5964 \pm 3.3737$      \\
$f_{\rm s, kmt}$                          &&    $0.0628 \pm 0.0030$        & $0.0773 \pm 0.0046$         &&   $0.0323 \pm 0.0069$         &   $0.0620 \pm 0.0066$       \\
$f_{\rm b, kmt}$                          &&    $0.1172 \pm 0.0026$        & $0.1022 \pm 0.0042$         &&   $0.1481 \pm 0.0070$         &   $0.1183 \pm 0.0064$       \\
\enddata
\tablecomments{ HJD$^\prime$ = HJD - 2450000.}
\end{deluxetable*}

\begin{figure*}
\centering
\includegraphics[width=0.8\textwidth]{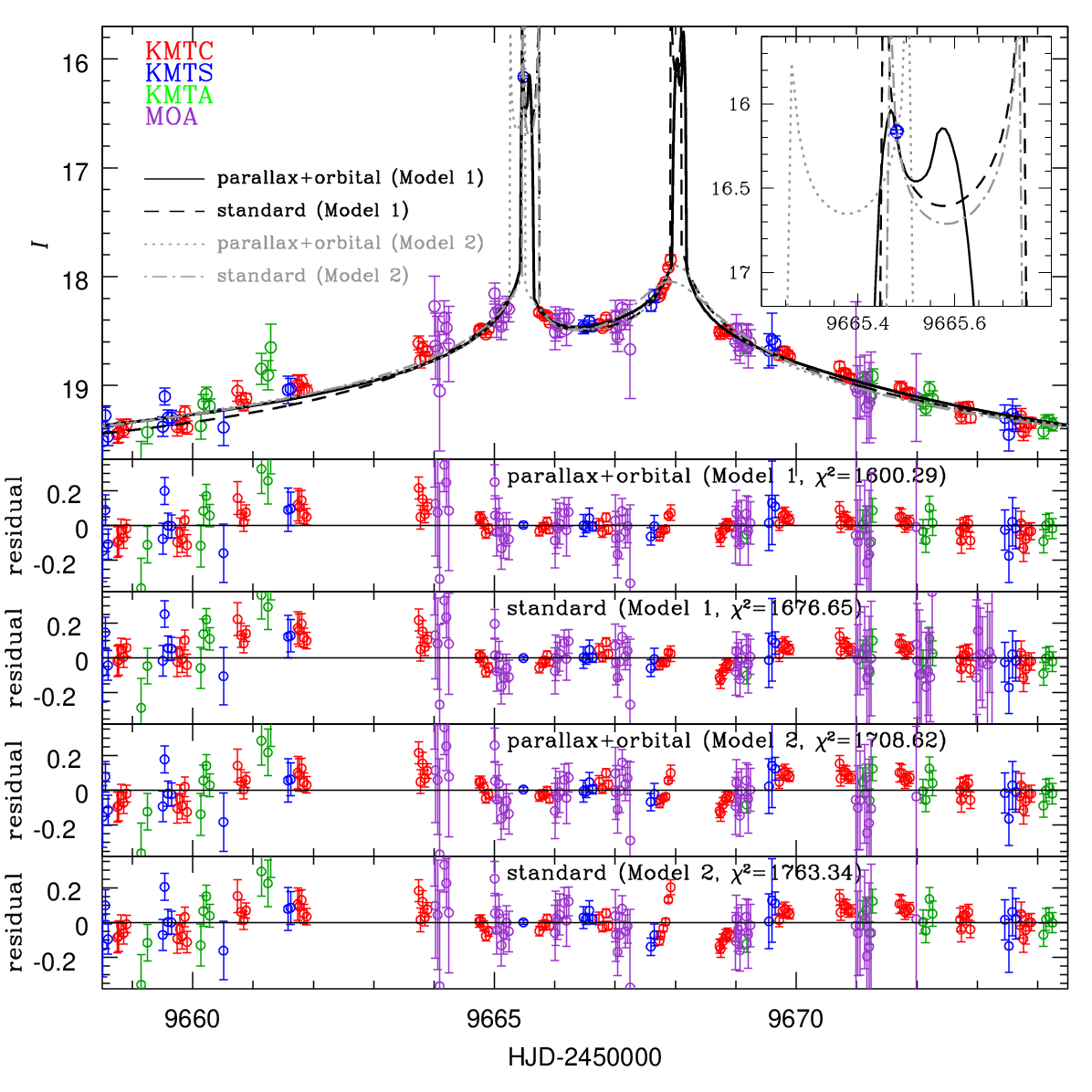}
\caption{Light curves of two different 2L1S models with and without high-order effects.
``Model 1" is the best-fit model and ``Model 2" is the alternative model with a different source trajectory.
The black solid and dashed curves represent the light curves of the ``Model 1" with and without high-order effects, respectively, while the gray dotted and dash-dotted curves represent the light curves for ``Model 2" .
The lower four panels present the residuals from the four models.
\label{fig:2L1S}}
\end{figure*}

\section{Light-curve Analysis}
Figure 1 presents the observed light curve of \thisevent, which has a U-shape feature in the peak that appears to be caused by caustic-crossing.
We thus conduct a standard binary-lens single-source (2L1S) modeling.

\begin{figure}
\centering
\includegraphics[width=0.5\textwidth]{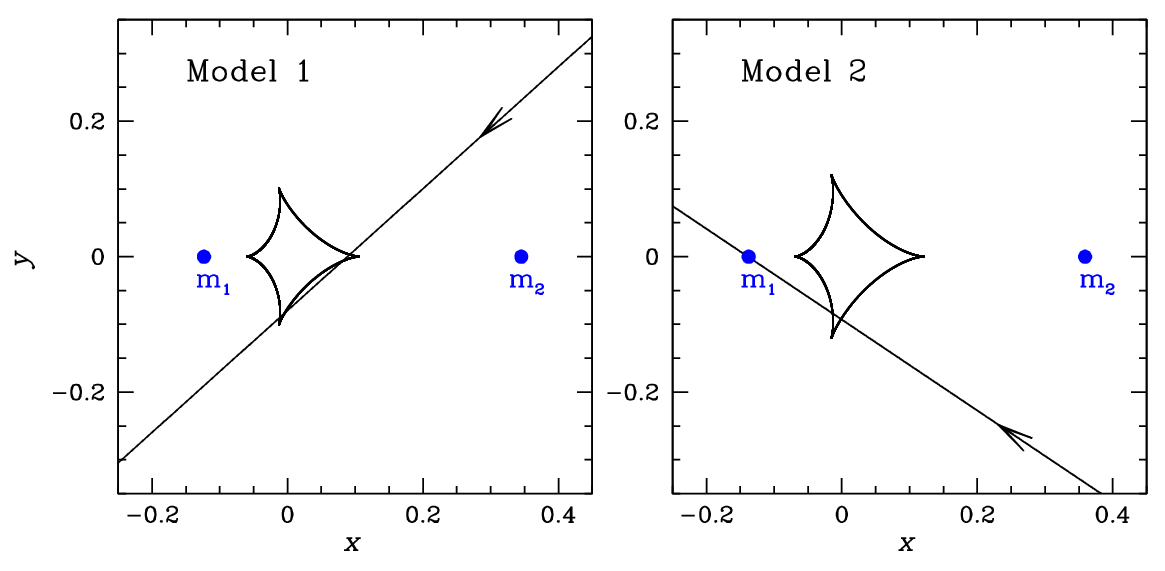}
\caption{Geometries of the two best-fit 2L1S models.
The blue dots represent the lens components of each model and the black closed curve represents the caustic.
The straight line with an arrow denotes the source trajectory.
\label{fig:2L1_geometry}}
\end{figure}

\subsection{2L1S model}
The standard 2L1S modeling requires seven parameters: three single lensing parameters $(t_0,u_0,\te)$, three binary lensing parameters $(s,q,\alpha)$, and $\rho$.
Here, $t_0$ is the time of the closest source approach to the lens, $u_0$ is the lens-source separation at $t=t_0$ in units of $\thetae$ (impact parameter), $\te$ is the Einstein radius crossing time of the event, $s$ is the projected separation of the binary lens components in units of $\thetae$, $q$ is the mass ratio of the two lens components, $\alpha$ is the angle between the binary axis and the source trajectory, and $\rho$ is the normalized source radius $\thetas/\thetae$, where $\thetas$ is the angular radius of the source.
In addition, there are two flux parameters $(f_{s,i},f_{b,i})$ for each observatory, which are the source flux and blended flux of the $i$th observatory, respectively.
The two flux parameters are modeled by $F_{i}(t) = f_{s,i}A_{i}(t) + f_{b,i}$, where $A_i$ is the magnification as a function of time at the $i$th observatory \citep{Rhie+1999}.

In order to find the binary lens solution, we first carry out a grid search in the parameter space $(s,q,\alpha)$ to find local $\chi^2$ minima using a downhill approach based on the Markov Chain Monte Carlo (MCMC) algorithm.
The ranges of each parameter used in the grid search are $-1 \leqslant \log s \leqslant 1$, $-4 \leqslant \log q \leqslant 0$, and $0 \leqslant \alpha \leqslant 2\pi$, and they are uniformly divided with $(50,50,50)$, respectively.
In the grid search, $s$ and $q$ are fixed, while the other parameters are allowed to vary in the MCMC chain.
From the grid search, we find two local solutions in the $(s,q)$ plane due to a well-known close-wide degeneracy \citep{Griest&Safizadeh1998}, in which they have similar $(s,q)$ but different trajectories of $\alpha \simeq 3.7$ and $2.4$.
We then refine the local solutions obtained from the grid search by allowing all parameters to vary.
From the procedures, it is found that the best 2L1S solution is the close model with $(s,q,\alpha)=(0.47,0.36,2.4)$ (Model 1), which approximately describes the anomalies, but there are significant residuals from the model.
This means that the standard model cannot explain the lensing light curve.
We note that the $\chi^2$ of the wide model is much larger than the best close model by $110$.
For the other solution with $\alpha=3.70$ (Model 2), it fails to describe the second anomaly around the peak, thus it is worse than the best solution by $\delcs=97$.
Here we note that Model 2 has a severe close-wide degeneracy and the close model is favored over the wide model by $\delcs=4$.
Figure 2 shows the light curves of the two binary lens models and residuals from the models.
Model 1 has two caustic-crossing features around the peak, while Model 2 has both caustic-crossing and cusp-approaching features (see Figure 3).
The lensing parameters of the two standard models are presented in Table \ref{fig:2L1S}.

The anomalies of the event can be caused by the high-order effects including the microlens parallax and lens orbital motion, which are caused by the orbital motions of the Earth and binary lens, respectively.
We thus conduct the binary lens modeling including the microlens parallax and lens orbital parameters that are described by $\bpie=(\pien, \piee)$ and $(ds/dt, d\alpha/dt)$.
In this modeling, we restrict the lens systems to $\beta <1$, where $\beta$ is a ratio of the projected kinetic to the potential energy, ${\rm (KE/PE)_\perp}$.
The parallax+orbital modeling is carried out based on the two 2L1S models.
As a result, it is found that the parallax+orbital model explains the light curve better than the standard model by $\delcs=86$, but it also cannot describe the light curve.
Figure 2 shows that there are still remarkable residuals from the models.
This indicates that it is highly likely that the event would be caused by an extra source (2L2S) or an extra lens (3L1S).

\begin{deluxetable*}{lcccccc}
\tablewidth{0pt}
\tablecaption{Best-fit lensing parameters of 2L2S models\label{tab-2L2S}}
\tablehead{
Parameter                                    &&                    \multicolumn{2}{c}{Model 1}                              &&          \multicolumn{2}{c}{Model 2}        \\
 \cline{2-7}                                 &&                Close                    &                Wide                        &&              Close                    &        Wide               
 }

\startdata
$\chi^2$                                  &&   $1513.42$                           & $1517.248$                            &&   $1493.68$                         & $1491.74$                        \\
$t_{01}$ (HJD$^\prime$)          &&   $9666.6661 \pm 0.0563$   & $9667.5836 \pm 0.0698$     &&  $9665.2832 \pm 0.0566$  & $9665.5305 \pm 0.0599$ \\
$t_{02}$ (HJD$^\prime$)          &&   $9665.5403 \pm 0.1298$   & $9665.3797 \pm 0.1765$     &&  $9668.0693 \pm 0.0712$  & $9668.2471 \pm 0.0390$ \\
$u_{01}$                                  &&  $0.0844 \pm 0.0032$          & $0.0829 \pm 0.0051$           &&   $0.0667 \pm 0.0109$        &  $-0.0340 \pm 0.0039$      \\
$u_{02}$                                  &&  $0.2329 \pm 0.0135$          & $-0.3788 \pm 0.0355$           && $-0.0964 \pm 0.0139$      & $0.0499 \pm 0.0058$       \\
$\te$ (days)                             &&  $12.7327 \pm 0.5302$        & $12.6105 \pm 0.7420$         &&  $15.9137 \pm 1.1454$      & $24.2254 \pm 2.8416$     \\
$q_F$                                      &&  $0.8718 \pm 0.1296$          & $2.4190 \pm 0.5403$           &&  $1.3410 \pm 0.0696$        & $1.2420 \pm 0.0540$       \\
$s$                                          &&  $0.5416 \pm 0.0094$          & $2.4874 \pm 0.0799$           &&   $0.4047 \pm 0.0232$       & $4.3619 \pm 0.2169$       \\
$q$                                         &&  $0.3673 \pm 0.0331$          & $0.5819 \pm 0.1398$            &&   $0.4229 \pm 0.0703$       & $1.9082 \pm 0.9697$       \\
$\alpha$ (radians)                   &&  $2.3757 \pm 0.0291$          & $2.6656 \pm 0.0337$            &&   $3.0356 \pm 0.0523$      & $-3.0565 \pm 0.0379$       \\
$\rho_1$                                 &&  $< 0.0013$                          & $<0.0013$                             &&   $<0.003$                         & $<0.002$       \\
$\rho_2$                                 &&  $<0.010$                               & -                                           &&  $<0.022$                          & $<0.014$       \\
$f_{\rm s, kmt}$                      &&    $0.1189 \pm 0.0068$        & $0.2347 \pm 0.0373$            &&  $0.0845 \pm 0.0108$       & $0.0842 \pm 0.0096$       \\
$f_{\rm b, kmt}$                      &&    $0.0627 \pm 0.0065$        & $-0.0541 \pm 0.0372$         && $0.0962 \pm 0.0104$       & $0.0963 \pm 0.0093$       \\
\enddata
\end{deluxetable*}

\begin{figure*}
\centering
\includegraphics[width=0.8\textwidth]{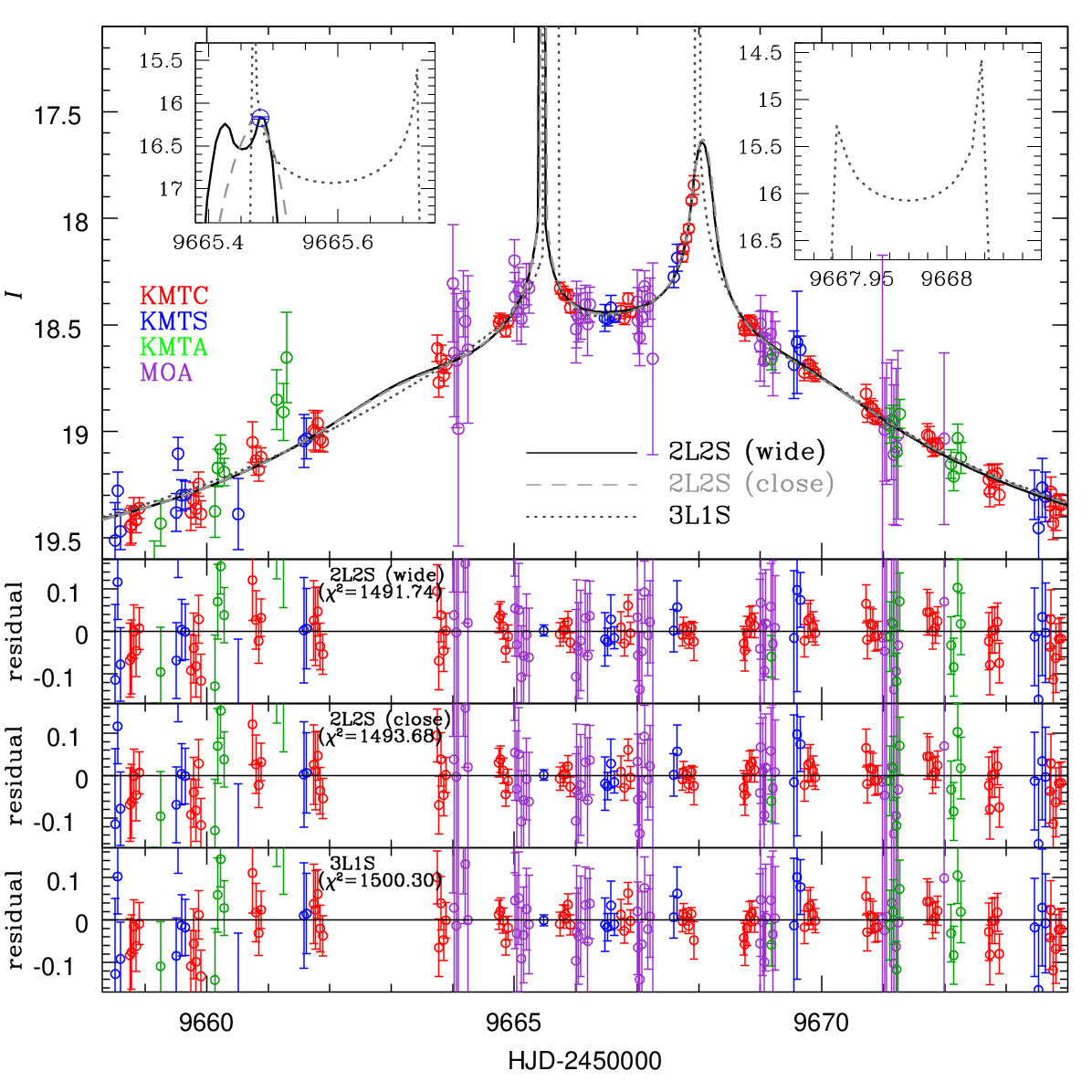}
\caption{Light curves of the best-fit 2L2S and 3L1S models.
The black solid and gray dashed curves represent the close and wide solutions of the 2L2S model, while the black dotted curve represents the light curve of the 3L1S model.
\label{fig:2L2S+3L1S}}
\end{figure*}

\subsection{2L2S model}
For the 2L2S modeling, four lensing parameters $(\tzerotwo,\uzerotwo,q_{\rm F},\rho_2)$ are added to those of the 2L1S model. 
Here, $\tzerotwo$ and $\uzerotwo$ are the peak time and impact parameter of the secondary source star ($S_2$), $q_{F}$ is the flux ratio between the secondary and primary source stars, and $\rho_2$ is the normalized angular radius of the $S_2$.
The subscripts ``1" and ``2" represent the lensing parameters related to the primary source ($S_1$) and secondary source, respectively.

2L2S modeling is generally conducted based on the best 2L1S solution.
For this event, there are two 2L1S solutions with different source trajectories of $\alpha=2.4$ and $\alpha=3.7$, although the $\delcs$ between the two models is 86.
Such a significant $\delcs$ was due to the $\alpha=3.7$ solution's failure to describe the second anomaly around the peak.
Since the second anomaly could be caused by the binary companion of the source, we carry out 2L2S modeling for the two 2L1S models.
In the 2L2S modeling, we use the lensing parameters of the 2L1S model to obtain initial values of the primary source, i.e., $(\tzerone,\uzerone,\te, s, q, \alpha, \rho_1)$, while we set the initial values of the secondary source, $(\tzerotwo, \uzerotwo, q_F)$, by considering the peak time and second anomaly magnification.
We then refine the initial parameters by allowing all parameters to vary.
Here we consider four degenerate solutions due to the unknown sign of $(\pm \uzerone,\pm \uzerotwo)$, i.e., the well-known ``ecliptic degeneracy" (\citealt{Jiang+2004, Poindexter+2005}).
We also consider the close and wide solutions for each 2L1S model.
From the modeling, we find that the best-fit 2L2S solution is the wide model based on ``Model 2", and it describes all the anomalies in the lensing light curve better than the best model based on ``Model 1" by $\delcs=22$.
The $\chi^2$ improvement mainly appears around the peak with the remarkable anomalies.
Therefore, we dismiss the 2L2S models based on ``Model 1".
The close and wide models for ``Model 2" have $\delcs < 2$.
Thus, they are severely degenerate.
Table \ref{tab-2L2S} presents the best-fit 2L2S lensing parameters, and their geometries are shown in Figure \ref{fig:geometry}.
As shown in Figure \ref{fig:2L2S+3L1S}, the 2L2S model describes well the anomaly range $9660 \leqslant {\rm HJD}^\prime \leqslant 9672$ that could not be explained by the 2L1S model.
The $\delcs$ improvement of the 2L2S model relative to the standard model is 185.

\begin{figure}[t!]
\centering
\includegraphics[width=0.5\textwidth]{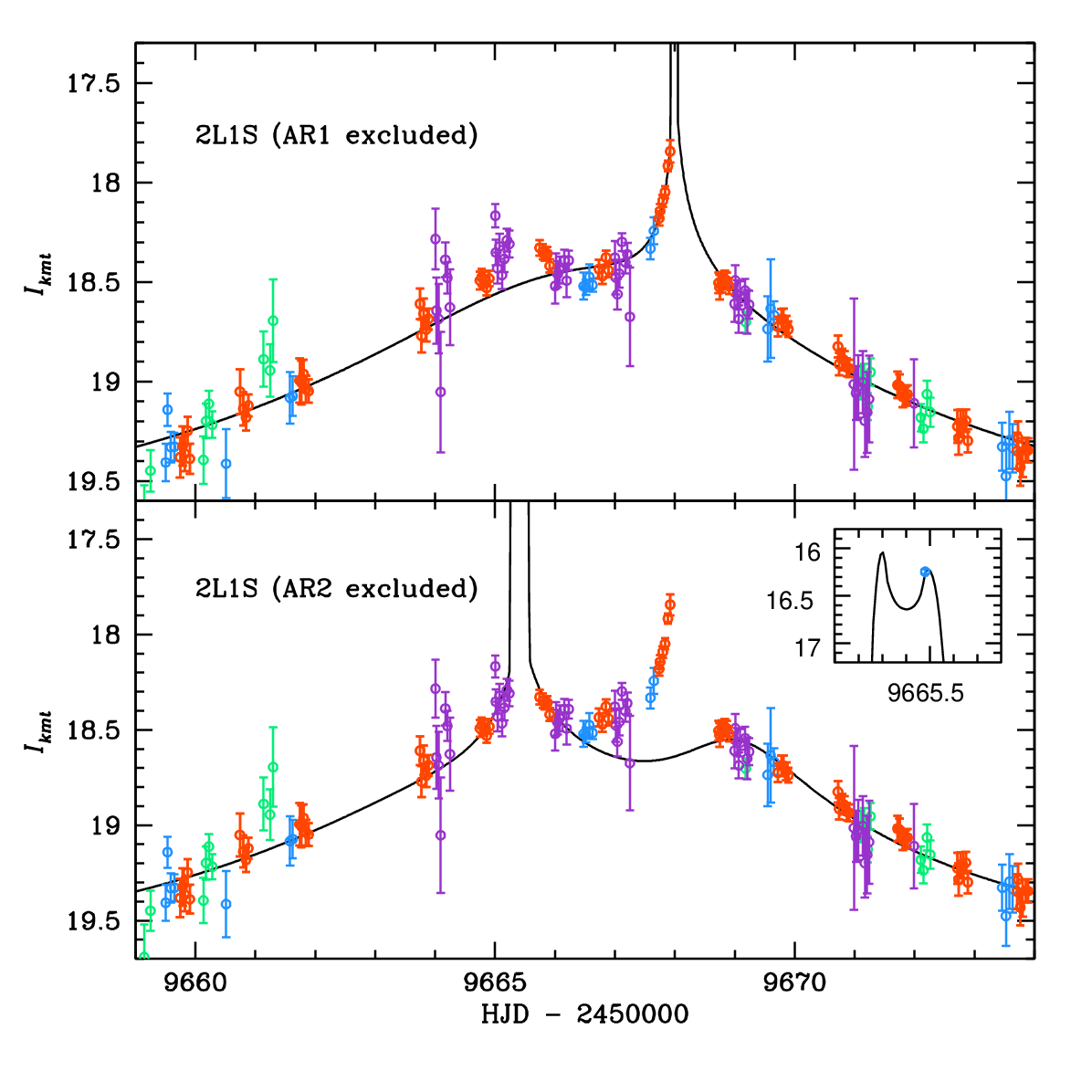}
\caption{Light curves of the 2L1S solutions where the anomaly regions 1 and 2 (AR1 and AR2) are individually excluded.
The AR1 and AR2 represent the ranges of  $9664 < {\rm HJD}^\prime < 9666.3$ and $9666.5 < {\rm HJD}^\prime < 9670.0$, respectively
\label{fig:two2L1S}}
\end{figure}

\begin{deluxetable*}{lcccccc}
\tablewidth{0pt}
\tablecaption{Best-fit lensing parameters of 2L1S model for AR1 or AR2 excluded\label{two2L1S}}
\tablehead{
                                                                   & \multicolumn{2}{c}{AR1 excluded}                       &&                     \multicolumn{2}{c}{AR2 excluded}             \\
\cline{2-6}                                              
Parameter                            &         $s<1$                             &            $s>1$                            &&               $s<1$                            &           $s>1$     
 }
\startdata
$t_0$ (HJD$^\prime$)        &  $9666.9960 \pm 0.0832$     &    $9667.1565 \pm 0.0464$      &&       $9666.4533 \pm 0.0502$       &    $9666.3440 \pm 0.0464$    \\
$u_0$                                & $0.2324 \pm 0.0332$             &   $0.1914 \pm 0.0345$             &&       $0.2546 \pm 0.0303$            &   $0.2654 \pm 0.0284$         \\
$\te$ (days)                       &  $9.8812 \pm 0.6568$            &   $12.2625 \pm 1.1579$           &&       $9.9918 \pm 0.6317$            &   $9.8712 \pm 0.6257$        \\
$s$                                    &  $0.8113 \pm 0.0485$            &   $2.1347 \pm 0.1331$             &&        $0.9294 \pm 0.0325$            &   $1.5721 \pm 0.0484$          \\
$q$                                   &  $0.1043 \pm 0.0276$            &   $0.3319 \pm 0.0705$             &&       $0.0327 \pm 0.0103$            &   $0.0793 \pm 0.0193$         \\
$\alpha$ (radians)            &  $1.1654 \pm 0.0203$            &   $1.2254 \pm 0.0220$              &&        $1.9435 \pm 0.0021$            &   $1.9238 \pm 0.0243$         \\
$\rho$                             &  $< 0.02$                                &   $< 0.015$                               &&         $< 0.011$                             &   $   < 0.006$        \\
\enddata
\tablecomments{The anomaly regions AR1 and AR2 represent the ranges of  $9664 < {\rm HJD}^\prime < 9666.3$ and $9666.5 < {\rm HJD}^\prime < 9670.0$, respectively.}
\end{deluxetable*}

\subsection{3L1S model}
The two major anomalies of \thisevent~appear around the peak, which means that they were caused by the caustic close to the primary star.
In this case, the anomalies induced by the triple lens systems can be approximated by the superposition of those induced by two binary lens systems (\citealt{Bozza1999, Han+2001, Han2005}).
We thus search for two binary lens models that can explain the overall light curve.
The two anomalies are individually in the ranges of $9664 < {\rm HJD}^\prime < 9666.3$ (anomaly region 1, hereafter AR1) and $9666.5 < {\rm HJD}^\prime < 9670.0$ (anomaly region 2, hereafter AR2).
After excluding each anomaly, we perform the same procedure as described in Section 3.1.
From this, we find that the best solution for the AR1 excluded is a binary star model with $q \sim 0.1-0.3$, while for the AR2 excluded it is likely a planetary model with $q \sim 0.03-0.08$.
This implies that the tertiary lens component would be a massive planet or a brown dwarf (BD).
The two best solutions each have a severely degenerate model with $\delcs < 1$ (for AR1 excluded) or $\delcs < 7$ (for AR2 excluded) due to the close-wide degeneracy.
Figure \ref{fig:two2L1S} shows the best-fit light curves for each case, and the resulting lensing parameters are presented in Table \ref{two2L1S}.

\begin{deluxetable*}{lcccccc}
\tablewidth{0pt}
\tablecaption{Best-fit lensing parameters of 3L1S model\label{tab-3L1S}}
\tablehead{
                                                                   & \multicolumn{2}{c}{$s_2 < 1$}                              &&                  \multicolumn{2}{c}{$s_2 >1$}  \\
\cline{2-6}                                              
Parameter                            &         Close-Close                     &            Close-Wide                    &&            Wide-Close                  &           Wide-Wide    \\
                                           &         $(s_3<1)$                        &       $(s_3>1)$                            &&            $(s_3<1)$                         &         $(s_3>1)$
 }
\startdata
$\chi^2$                             &   $1514.17$                             &         $1523.09$                        &&       $1500.30$                             &         $1517.88$                    \\
$t_0$ (HJD$^\prime$)         &  $9666.6985 \pm 0.0497$     &    $9666.6381 \pm 0.0649$      &&       $9666.9538 \pm 0.0414$     &    $9666.9663 \pm 0.0470$    \\
$u_0$                                 & $0.2765 \pm 0.0241$             &   $0.2581 \pm 0.0300$             &&       $0.2369 \pm 0.0203$             &   $0.1900 \pm 0.0273$         \\
$\te$ (days)                        &  $9.5319 \pm 0.5523$          &   $10.1363 \pm 0.6729$           &&       $10.4896 \pm 0.5886$          &   $12.8868 \pm 1.0883$        \\
$s_2$                                 &  $0.9172 \pm 0.0323$            &   $0.8920 \pm 0.0383$             &&       $1.7224 \pm 0.0635$            &   $1.8765 \pm 0.0890$          \\
$q_2$                                 &  $0.0221 \pm 0.0052$            &   $0.0239 \pm 0.0093$             &&       $0.1320 \pm 0.0247$            &   $0.1825 \pm 0.0304$         \\
$\alpha$ (radians)              &  $1.0202 \pm 0.0214$            &   $0.9445\pm 0.0216$             &&       $1.0566 \pm 0.0168$            &   $1.0268 \pm 0.0203$         \\
$s_3$                                 &  $0.9410 \pm 0.0256$            &   $1.5220 \pm 0.0328$             &&       $0.9031 \pm 0.0200$            &   $1.3983 \pm 0.0365$         \\
$q_3$                                 & $0.0381\pm 0.0065$             &   $0.0635 \pm 0.0117$              &&       $0.0465\pm 0.0057$             &   $0.0556 \pm 0.0104$         \\
$\psi$ (radians)                  &  $0.9675 \pm 0.0367$            &   $1.0617 \pm 0.0452$             &&       $1.1104 \pm 0.0333$            &   $1.2017 \pm 0.0365$         \\
$\rho$                               &  $< 0.007$                              &    $ < 0.006$                              &&     $< 0.003$                                &   $ < 0.004$        \\
$f_{\rm s, kmt}$                 &  $0.2323 \pm 0.0239$            &   $0.2208 \pm 0.0316$             &&       $0.2211 \pm 0.0199$            &   $0.1743 \pm 0.0326$         \\
$f_{\rm b, kmt}$                 & $-0.0504 \pm 0.0236$           &   $-0.0393 \pm 0.0313$           &&      $-0.0397 \pm 0.0197$           &   $0.0061 \pm 0.0323$        \\
\enddata
\end{deluxetable*}

\begin{figure}[t!]
\centering
\includegraphics[width=0.5\textwidth]{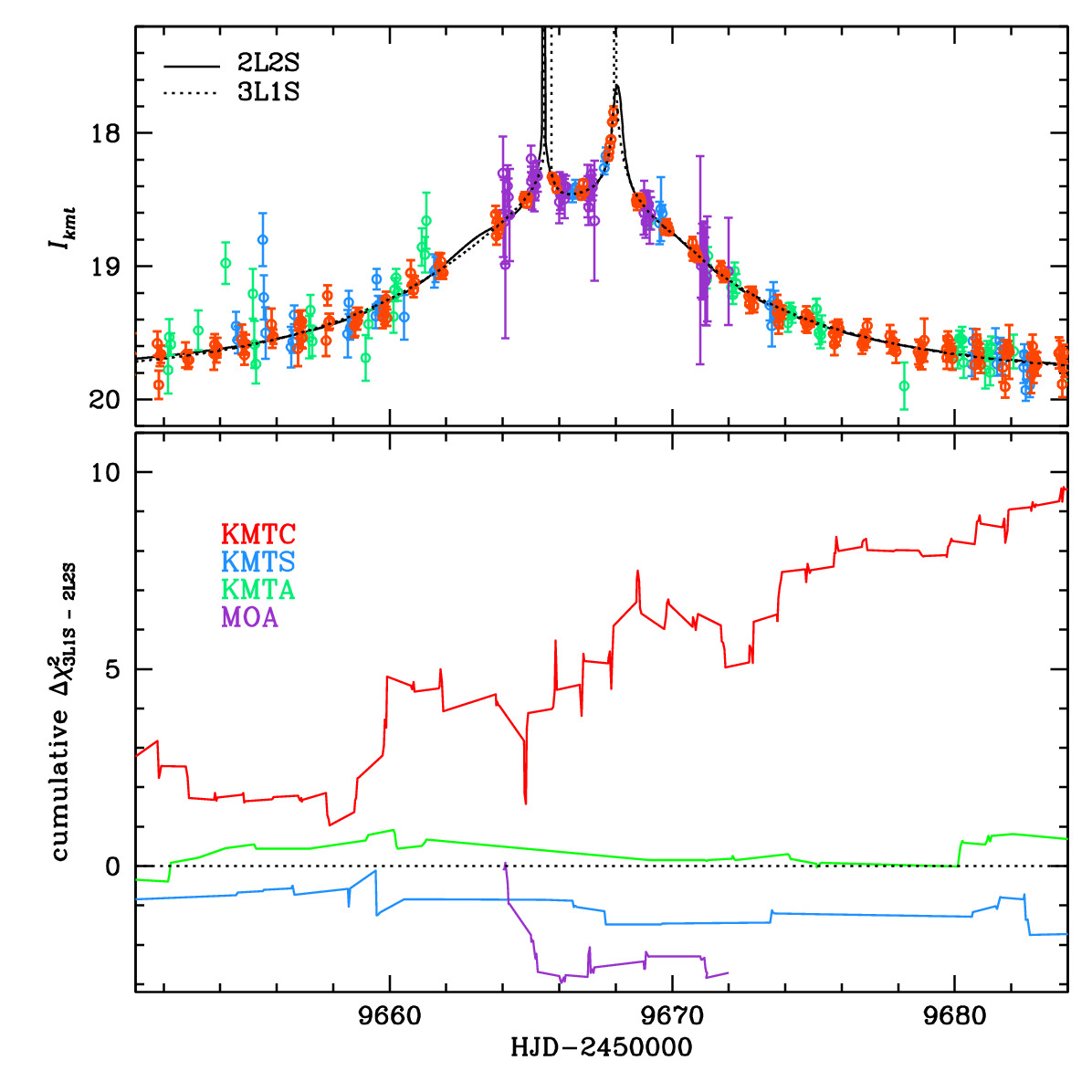}
\caption{Cumulative $\delcs$ distribution between the 3L1S and 2L2S models.
\label{fig:cumulative}}
\end{figure}

\begin{figure}[t!]
\centering
\includegraphics[width=0.5\textwidth]{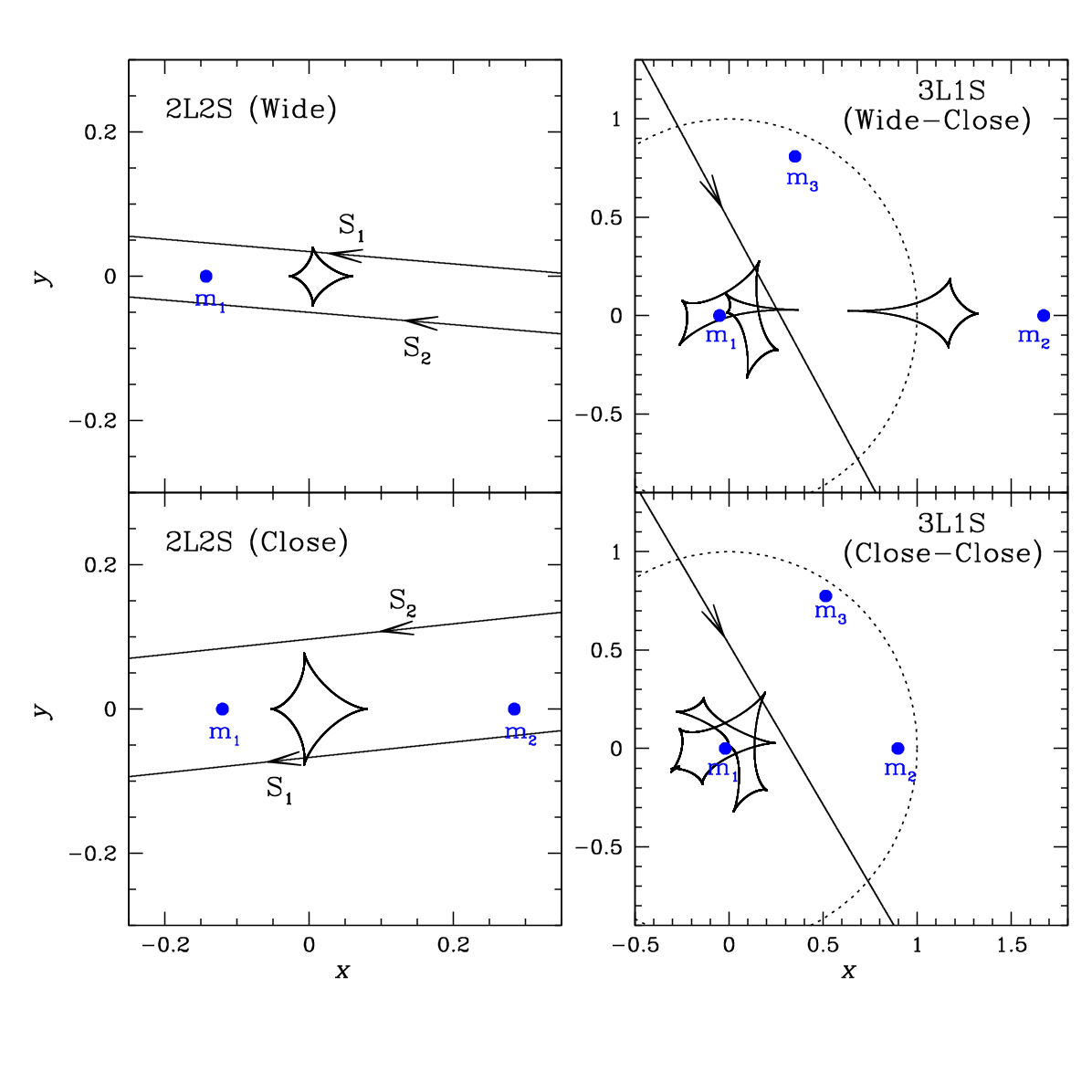}
\caption{Geometries of the best-fit 2L2S and 3L1S models.
For the 2L2S model, $S_{1}$ and $S_{2}$ denote the primary and secondary sources.
The dotted circle represents the Einstein ring and the other notations are the same as Figure 3.
\label{fig:geometry}}
\end{figure}

For the triple lens modeling, three parameters related to an extra lens component are added, and they are $(s_3,q_3,\psi)$, where $s_3$ is the separation between the primary and the tertiary components, $q_3$ is the tertiary-primary mass ratio ($M_3/M_1$), and $\psi$ is the angle between the primary-secondary and the primary-tertiary axes.
In conformity with the notation of the subscript ``3" for the primary-tertiary parameters, we set the primary-secondary parameters to the subscript ``2", that is, $(s_2,q_2)$.
In order to find the 3L1S solution, we perform a grid search in the parameter space $(s_3,q_3,\psi)$.
The parameters $(s_3,q_3,\psi)$ have the ranges of $-1 \leqslant \log s_3 \leqslant 1$, $-4 \leqslant \log q_3 \leqslant 0$, and $0 \leqslant \psi \leqslant 2\pi$ and they are uniformly divided with $(50,50,50)$.
In this grid search, $(s_2,q_2,\alpha)$ are fixed, while the other lensing parameters are allowed to vary.
The initial lensing parameters are set to the close ($s_2<1$) and wide ($s_2>1$) solutions for the AR1 excluded (i.e., binary star model), respectively.
Each grid search produces two locals that represent the close and wide solutions.
We then refine the local solutions by allowing all parameters to vary.
As a result, we find that the best-fit 3L1S solution is $(s_2,q_2,s_3,q_3)=(1.72,0.13,0.90,0.05)$.
Here we note that the solution for $s_2 > 1$ and $s_3 < 1$ is denoted as the ``Wide-Close" model, while for $s_2 < 1$ and $s_3<1$ it is denoted as the ``Close-Close" model.
The ``Close-Close" model is disfavored by $\delcs=14$ relative to ``Wide-Close" model.
The lensing parameters of the best-fit 3L1S model, ``Wide-Close" model, is presented in Table \ref{tab-3L1S} with the other three models, and its light curve is shown in Figure \ref{fig:2L2S+3L1S} together with those of the 2L2S models.
Figure \ref{fig:2L2S+3L1S} shows that the triple lens model also describes the light curve well.
However, the 2L2S model is preferred over the triple lens model by $\delcs=8.6$.
Figure \ref{fig:cumulative} is the cumulative $\delcs$ distribution between the 3L1S and 2L2S models and shows that the $\chi^2$ improvement is due to dominant KMTC data sets of the light curve that provide a better fit throughout the light curve, including the anomaly regions.
This implies that \thisevent~is likely caused by the 2L2S model.

We also check the 3L1S models based on the two 2L1S models, i.e., ``Model 1" and ``Model 2".
All the 3L1S models based on the two models have still severe residuals from each model.
The 3L1S models for the ``Model 1" and ``Model 2" are worse than the best-fit 3L1S model by $\delcs=79$ and 106, respectively.
Hence, they also cannot explain the light curve of the event.

\begin{figure}[t!]
\centering
\includegraphics[width=0.5\textwidth]{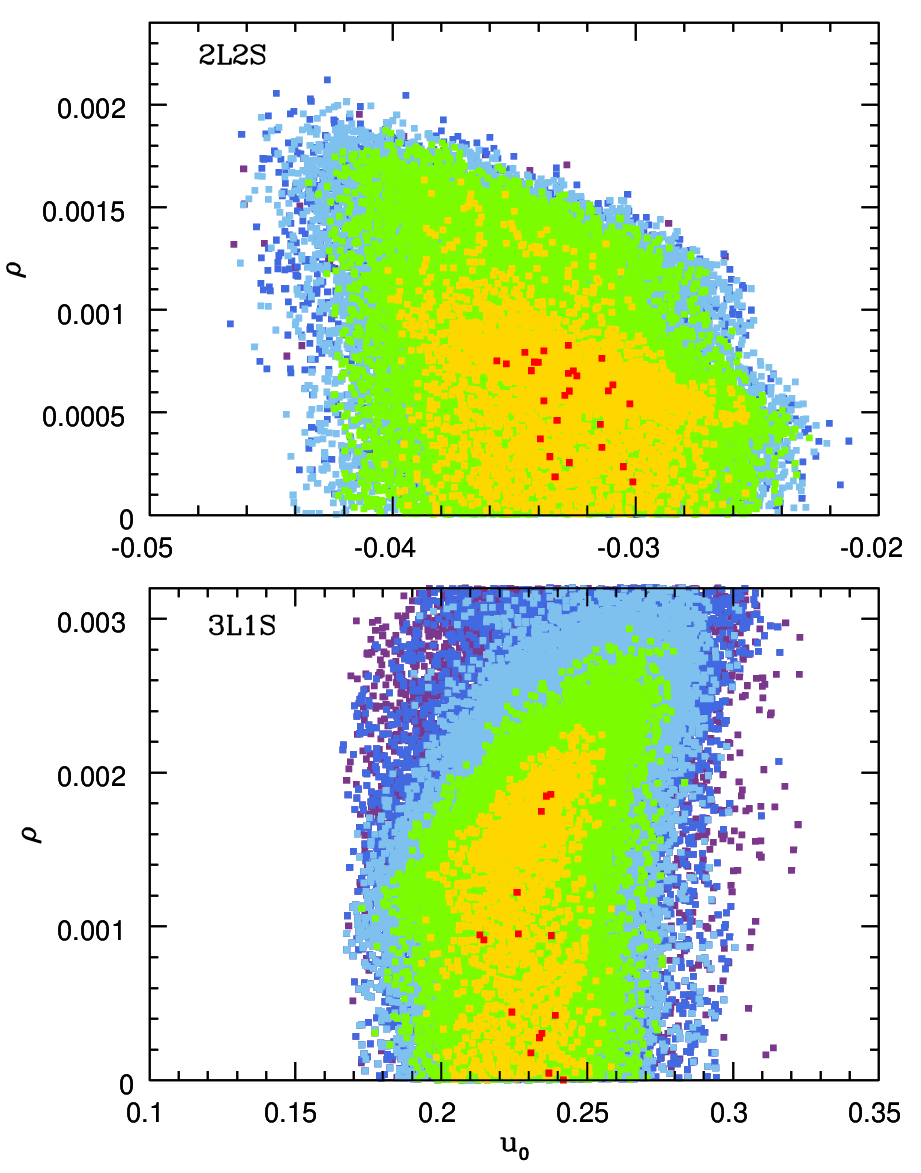}
\caption{$\chi^2$ distributions of $\uzero$ and $\rho$ for the best-fit 2L2S and 3L1S models.
The $\rho$ of the 2L2S model represents the value of the primary source.
\label{fig:rho}}
\end{figure}

\section{Angular Einstein radius}

The angular Einstein radius is one of two key parameters to measure the physical lens parameters including the lens mass ($M_{\rm L}$) and distance to the lens ($D_{\rm L}$) and is defined by $\thetae = \theta_\star/\rho$.
For the event, the two major anomalies around the peak were not well covered.
Thus, we only measured the upper limit of $\rho$, i.e., $\rho_{\rm max,s1} \simeq 0.002$ within the $3\sigma$ level (see Figure \ref{fig:rho}).
We thus can only obtain the lower limit of the $\thetae$.

\begin{figure}[t!]
\centering
\includegraphics[width=0.5\textwidth]{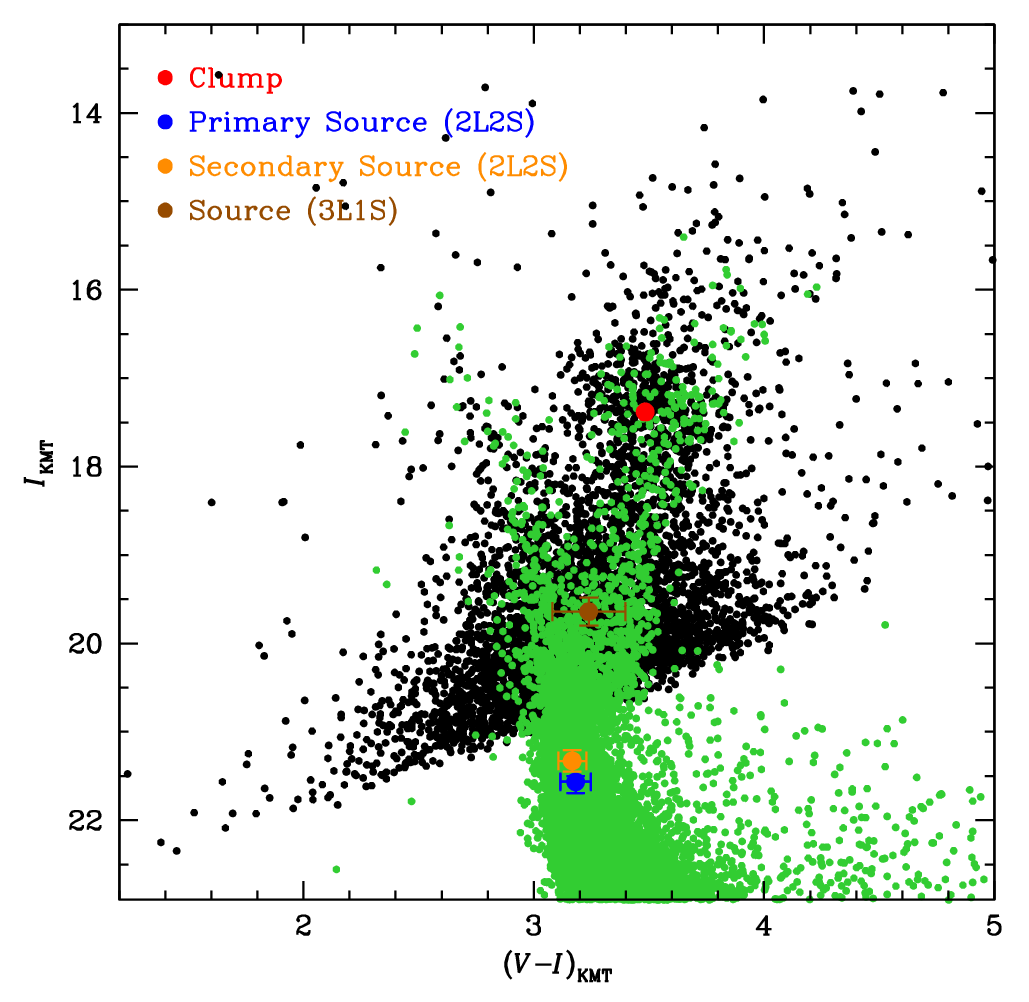}
\caption{Instrumental color-magnitude diagram (CMD) of stars in the observed field, which is constructed from combining KMTC and \textit{HST} observations.
The KMTC and \textit{HST} CMDs are plotted as black and green dots, respectively.
The blue and cyan dots represent the positions of the primary and secondary sources for the 2L2S model, and the brown dot represents the source position for the 3L1S model. The red dot denotes the red giant clump centroid.
\label{fig:cmd}}
\end{figure}

The angular source radius is obtained from the de-reddened color and magnitude of the source that are determined from the offset between the source star and the red giant clump positions on the CMD, 
\begin{equation}
(V-I, I)_{\rm s,0} = (V-I, I)_{\rm cl,0} + \Delta(V-I,I),
\end{equation}
where $\Delta(V-I,I) = (V-I,I)_{\rm s} - (V-I,I)_{\rm cl}$ represents the offset between the source and the centroid of the red giant clump.
This is based on the assumption that the source and clump experience the same amount of reddening and extinction \citep{Yoo+2004}.
From the KMTC CMD constructed from stars around the source star, we find that the centroid of the clump is $(V-I, I)_{\rm cl}=(3.48 \pm 0.17, 17.38 \pm 0.43)$.
The measured magnitudes of the primary and secondary source stars are $I_{\rm s1}=21.56 \pm 0.13$ and $I_{\rm s2}=21.33 \pm 0.12$, which are obtained from the source fluxes of the best-fit 2L2S wide model.
Due to only two low-magnified KMTC $V$-band data points, the source color was not securely measured.
We thus combine the KMCT CMD and the CMD constructed by \textit{Hubble Space Telescope} (\textit{HST}) toward Baade's window \citep{Holtzman+1998}.
The combination of the two CMDs is conducted by calibrating the clump positions on the individual CMDs.
Figure \ref{fig:cmd} shows the combined CMD.
The source color is estimated from the mean value of the \textit{HST} stars with similar magnitudes to the source star.
From the combined CMD, we obtain the primary and secondary source colors of $(V-I)_{\rm s1}=3.182 \pm 0.066$ and $(V-I)_{\rm s2}=3.167 \pm 0.061$.
The de-reddened color and magnitude of the clump are adopted by $(V-I)_{\rm cl,0}=1.06$ and $I_{\rm cl,0}=14.43$ from \citet{Bensby+2011} and \citet{Nataf+2013}, respectively.
As a result, it is found that the de-reddened colors and magnitudes of the two sources are $(V-I, I)_{\rm s1,0} =(0.758 \pm 0.066, 18.62 \pm 0.13)$ and $(V-I, I)_{\rm s2,0} =(0.743 \pm 0.061, 18.38 \pm 0.12)$.
This means that both stars are late G dwarf stars.
Using the $VIK$ color-color relation of \citet{Bessell&Brett1988} and color-surface brightness of \citet{Kervella+2004}, we obtain the angular radius of the primary source of  $\theta_{\star,\rm s1} = 0.615 \pm 0.056\, \rm \mu as$.
We then determine the lower limit of $\thetae$,
\begin{displaymath}
\theta_{\rm E,min} =  {\theta_{\star,\rm s1}\over{\rho_{\rm max,s1}}} = \left\lbrace
\begin{array}{ll}
0.209\, {\rm mas} & \textrm{for close} \\
0.308\, {\rm mas} & \textrm{for wide}.
\end{array}\right.
\end{displaymath}
The lower limit of the relative lens-source proper motion is obtained as
\begin{displaymath}
\mu_{\rm rel,min} = {\theta_{\rm E,min}\over{\te}} = \left\lbrace
\begin{array}{ll}
4.79\, {\rm mas\, yr^{-1}} & \textrm{for close} \\
4.64\, {\rm mas\, yr^{-1}} & \textrm{for wide}.
\end{array}\right.
\end{displaymath}

For the 3L1S model, by following the same procedure as the binary source model, we find that the de-reddened color and magnitude of the source are $(V-I,I)_0=(0.81 \pm 0.16, 16.69 \pm 0.10)$, indicating that the source is an early G-type subgiant or a turn-off star.
The color and magnitude derive the angular source radius of $\theta_\star=1.63 \pm 0.31\, \mu as$.
For this model, the upper limit of the normalized source radius is $\rho_{\rm max} \simeq 0.003$.
With the upper limit, we estimate the lower limits of the angular Einstein radius and relative lens-source proper motion, $\theta_{\rm E,min} = 0.542\, \rm mas$ and $\mu_{\rm rel,min}=18.89\, {\rm mas\, yr^{-1}}$, respectively.

\begin{figure}[t!]
\centering
\includegraphics[width=0.5\textwidth]{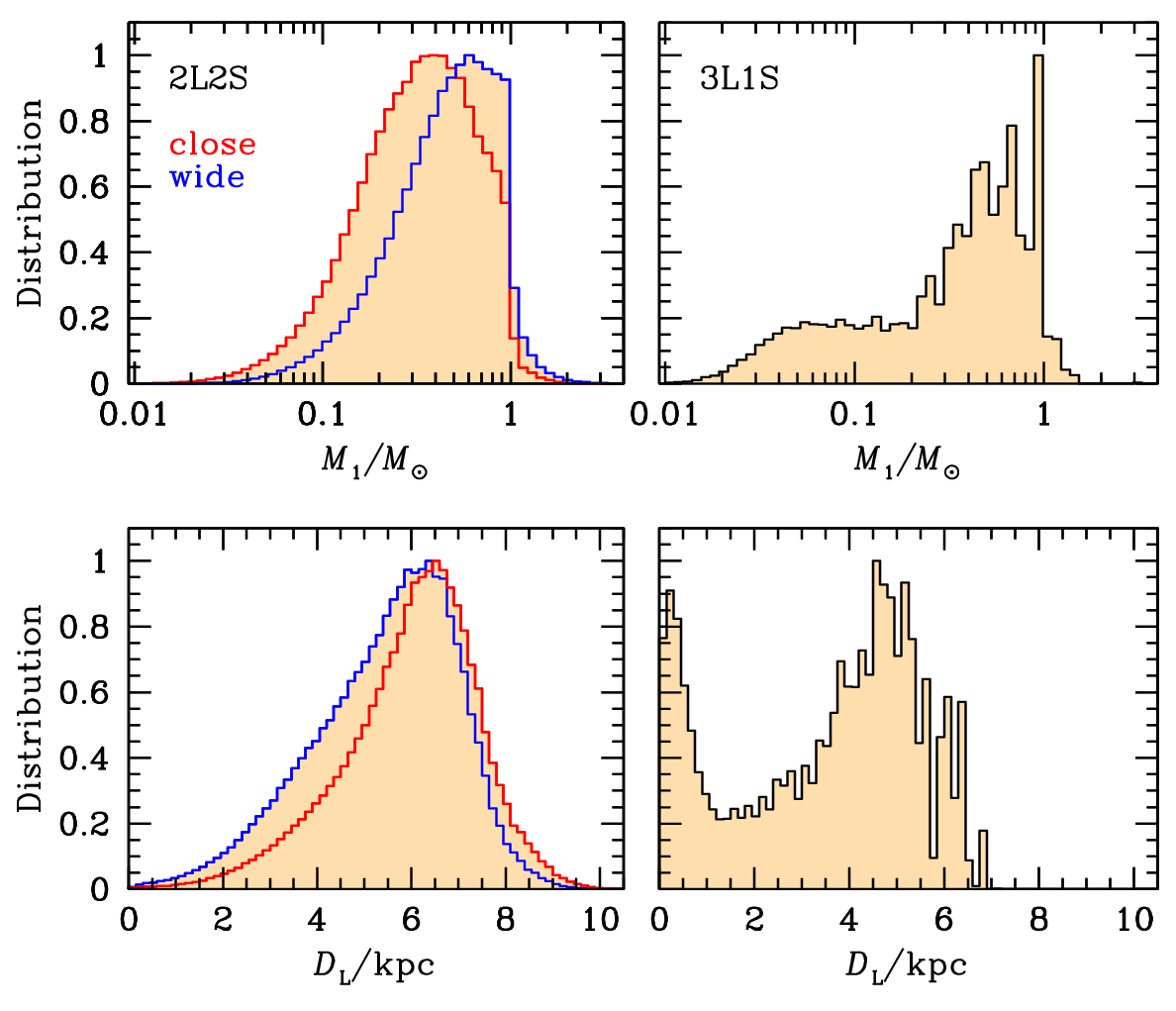}
\caption{Bayesian posteriors for the mass and distance of the host lens star for the 2L2S and 3L1S models.
The red and blue lines for the 2L2S model represent the distributions of the close and wide solutions.
\label{fig:Bayesian}}
\end{figure}

\section{Lens properties}

The lens mass and distance to the lens are given by
\begin{equation}
\label{eqn:mass}
M_{\rm L} =  {\thetae\over{\kappa \pie}};\quad \dl = {{\rm au}\over{\pie\thetae + \pi_{\rm S}}},
\end{equation}
where $\kappa \equiv 4G/(c^{2}\rm au) \approx 8.14\,{\rm mas}\, {M_\odot}^{-1}$ and $\pi_{\rm S}={\rm au}/\ds$ denotes the parallax of the source.
Hence, we can directly determine the physical lens parameters by measuring $\thetae$ and $\pie$, but it is generally hard to measure them.
For this event, only the lower limit of $\thetae$ was measured.
We thus carry out a Bayesian analysis to estimate the physical lens parameters with the measured two parameters $(\te,\theta_{\rm E,min})$.
The Bayesian analysis is conducted by using the Galactic model proposed by \citet{Jung+2021}.

\begin{deluxetable*}{lcccccccccc}
\tablewidth{0pt}
\tablecaption{Physical lens parameters\label{tab-lens}}
\tablehead{
                                                        &&                                                      \multicolumn{6}{c}{Physical parameters}                                                                                                                   &&    \multicolumn{2}{c}{Relative weights}   \\
\cline{2-11}                                              
      Models     &&   $M_{\rm 1}$ $(M_\odot)$  &    $M_{\rm 2}\, (M_\odot)$ &     $M_{\rm p}\, (M_{\rm J})$  &     $D_{\rm L}$ (kpc)         &       $a_{\perp}$ (au)          &   $a_{\perp,3}$ (au)      &&   Gal. Mod. & $\chi^2$        
 }
\startdata
2L2S                   \\
Close            &&  $0.35^{+0.34}_{-0.19}$     &  $0.15^{+0.15}_{-0.09}$   &                                      &  ${6.24^{+1.13}_{-1.62}}$  &  $0.53^{+0.58}_{-0.59}$  &          &&   0.90      &    0.38 \\
Wide             &&  $0.50^{+0.36}_{-0.27}$     &  $0.96^{+0.85}_{-0.71}$   &                                    &  ${5.75^{+1.24}_{-1.85}}$  &  $7.71^{+5.77}_{-6.06}$  &                               &&   1.00      &    1.00 \\
Adopted       &&  $0.46^{+0.35}_{-0.25}$     &  $0.75^{+0.67}_{-0.55}$   &                                       &  $5.87^{+1.21}_{-1.79}$    &   (bi-modal)                        &         &&                 & \\
\cline{1-11}
3L1S              &&  $0.43^{+0.41}_{-0.35}$     &  ${0.056^{+0.055}_{-0.047}}$   &  $20.84^{+20.20}_{-17.04}$ &  ${4.06^{+1.39}_{-3.28}}$  &  $3.79^{+3.72}_{-2.48}$  & $1.99^{+1.95}_{-1.30}$ &&  0.004    &    0.01 \\
\enddata
\end{deluxetable*}

We randomly create about $2\times10^7$ simulated microlensing events.
Then, we weight each simulated event, $i$, by
\begin{equation}
w_i = {\rm exp} \left( -{\chi^2_i\over{2}} \right);\quad  \chi^2_i = \left[ {t_{{\rm E},i}-\te \over{\sigma(\te)}}\right]^2,
\end{equation}
where $\sigma(\te)$ is the uncertainty of the measured value.
We set $w_i = 0$ if $\theta_{{\rm E},i} \leqslant \theta_{\rm E,min}$.
Figure \ref{fig:Bayesian} shows the Bayesian posteriors for the physical parameters of the primary lens star for close and wide solutions.
The estimated mass and distance of the primary star are
\begin{displaymath}
M_1 = \left\lbrace
\begin{array}{ll}
0.35^{+0.34}_{-0.19}\, M_\odot &  \textrm{for close} \\
0.50^{+0.36}_{-0.27}\, M_\odot &  \textrm{for wide},
\end{array} \right.
\end{displaymath}
and
\begin{displaymath}
\dl = \left\lbrace
\begin{array}{ll}
6.24^{+1.13}_{-1.62}\, {\rm kpc} &  \textrm{for close} \\
5.75^{+1.24}_{-1.85}\, {\rm kpc} &  \textrm{for wide}.
\end{array} \right.
\end{displaymath}
Then, the mass of the secondary component is
\begin{displaymath}
M_2 = \left\lbrace
\begin{array}{ll}
0.15^{+0.15}_{-0.09}\, M_\odot &  \textrm{for close} \\
0.96^{+0.85}_{-0.71}\, M_\odot &  \textrm{for wide}.
\end{array} \right.
\end{displaymath}
The projected separation of the lens components is
\begin{displaymath}
 a_{\perp} = \left\lbrace
\begin{array}{ll}
0.53^{+0.58}_{-0.59}\, {\rm au} &  \textrm{for close} \\
7.71^{+5.77}_{-6.06}\, {\rm au} &  \textrm{for wide}.
\end{array} \right.
\end{displaymath}
The results indicate that \thisevent L is likely to be a binary star composed of M and G dwarfs or two M dwarfs.
However, the primary star could be also a K dwarf and the secondary companion could be also a BD, a K, or an A dwarf.
Figure \ref{fig:Bayesian} shows that the lens system for the close and wide solutions is located in the disk and bulge with equal probability.
This is consistent with the relative lens-source proper motion of $\murel \geqslant 4.6\, {\rm mas\, yr^{-1}}$.

We also carry out the Bayesian analysis for the 3L1S model.
From the Bayesian analysis, it is found that the lens is a low-mass dwarf binary of $(M_1, M_2)=(0.43^{+0.41}_{-0.35}\, M_\odot, 0.056^{+0.055}_{-0.047}\, M_\odot)$ hosting a BD or a massive giant planet with a mass of $M_3=20.84^{+20.20}_{-17.04}\, M_{\rm J}$.
The result is shown in Figure 10.
The projected separations of the individual components from the primary star are $a_{\perp,2}=3.79^{+3.72}_{-2.48}\, \rm au$ and $a_{\perp,3}=1.99^{+1.95}_{-1.30}\, \rm au$, respectively.
If the tertiary component is a giant planet, it orbits beyond the snow line of the primary star.
The lens system is located at a distance of $\dl = 4.06^{+1.39}_{-3.28}\, \rm kpc$, which indicates that the lens is located in the disk of our Galaxy.
The proper motion of the triple lens model of $\murel \gtrsim 18.9\, {\rm mas\, yr^{-1}}$ is more than three times larger than a typical bulge-source disk-lens proper motion of $\sim 6\, \rm mas\, yr^{-1}$, so it does not seem to be a valid solution.

In order to check the unlikeness of the 3L1S solution, we estimate the relative weights of the 3L1S solution \citep{Gould+2022}.
The weights are the product of the two factors: the first is simply the total weight from the Bayesian analysis and the second is $\rm exp{(-\delcs/2)}$, where $\delcs$ is the $\chi^2$ difference relative to the best solution.
The two factors are presented at the right side of each row in Table 5.
For the 3L1S model, the relative weight is 0.00004, indicating that this solution is extremely unlikely.

We also estimate the relative weights of the close and wide solutions for the 2L2S model.
With the relative weights, we combine the physical parameters of the close and wide solutions and provide a single ``adopted" value for each parameter, which is simply the weighted average of the two solutions.
For $a_\perp$, we follow a similar approach, provided that either the individual solutions are strongly overlapping or one solution is strongly dominant.
If neither condition is met, we enter ``bi-model" instead.
The ``adopted" values are presented in Table 5.

\section{Summary}
We analyzed the microlensing event \thisevent~with remarkable anomalies around the peak.
From the analysis, it was found that the event was best described by the 2L2S model, but the 3L1S model is only disfavored by $\delcs\simeq 9$.
Unfortunately, the anomalies  were not enough covered, we thus could only measure the minimum angular Einstein radius $\theta_{\rm E,min}$.
From the Bayesian analysis with the measured $\te$ and $\theta_{\rm E,min}$, it was estimated that the masses of the binary lens components are $(M_1, M_2)=(0.46^{+0.35}_{-0.25}\, M_\odot, 0.75^{+0.67}_{-0.55}\, M_\odot)$, while those of the triple lens components are $(M_1, M_2, M_3)=(0.43^{+0.41}_{-0.35}\, M_\odot, 0.06^{+0.06}_{-0.05}\, M_\odot, 20.84^{+20.20}_{-17.04}\, M_{\rm J})$.
The relative lens-source proper motion of the 2L2S model of $\murel \geqslant 4.6\, {\rm mas\, yr^{-1}}$ corresponds to the Bayesian result with $\dl = 5.87^{+1.21}_{-1.79}$ kpc, whereas that of the 3L1S model of $\murel \geqslant 18.9\, {\rm mas\, yr^{-1}}$ does not correspond to the Bayesian result with $\dl = 4.06^{+1.39}_{-3.28}\, \rm kpc$, because it is $> 3$ times larger than the typical disk object value.
Therefore, it is likely that the event was caused by the 2L2S model.

\section{Acknowlegement}
The work by S.-J.C. was supported by the Korea Astronomy and Space Science Institute under the R\&D program (project No. 2025-1-830-05) supervised by the Ministry of Science and ICT.
The MOA project is supported by JSPS KAKENHI Grant Number JP24253004, JP26247023, JP16H06287, and JP22H00153.
J.C.Y., I.-G.S. acknowledge support from N.S.F grant No. AST-2108414.
W.Zang and H.Y. acknowledge support by the National Natural Science Foundation of China (Grant No. 12133005). 
Y.S. acknowledges support from BSF Grant No. 2020740.
This research has made use of the KMTNet system operated by the KASI and the data were obtained at three sites of CTIO in Chile, SAAO in South Africa, and SSO in Australia. 
Data transfer from the host site to KASI was supported by the Korea Research Environment Open NETwork (KREONET).

\end{document}